\documentclass[11pt,a4paper]{article}

\usepackage{amssymb}
\usepackage{graphicx}
\usepackage{bm}

\usepackage{mathrsfs}
\usepackage{cite}

\newcommand*\xbar[1]{%
  \hbox{%
    \vbox{%
      \hrule height 0.5pt 
      \kern0.5ex
      \hbox{%
        \kern-0.1em
        \ensuremath{#1}%
        \kern-0.1em
      }%
    }%
  }%
}

\begin{document}

\title{Exact relations between running of $\alpha_s$ and $\alpha$\\ in ${\cal N}=1$ SQCD+SQED}

\author{
A.L.Kataev${}^{a,b}$ and K.V.Stepanyantz${}^c$\\
\\
${}^a${\small{\em Institute for Nuclear Research of the Russian Academy of Science,}}\\
{\small {\em 117312, Moscow, Russia}},\\
${}^b${\small{\em Bogoliubov Laboratory of Theoretical Physics,}}\\
{\small{\em Joint  Institute for Nuclear Research ,}}\\
{\small {\em 141980, Dubna, Russia}},\\
${}^c${\small{\em Moscow State University}}, {\small{\em  Physical
Faculty, Department  of Theoretical Physics}}\\
{\small{\em 119991, Moscow, Russia}}\\
}

\maketitle

\begin{abstract}

In ${\cal N}=1$ supersymmetric QCD-like theories we derive the (all-order) exact equations relating the renormalization group behaviour of the strong and electromagnetic couplings and prove that they are valid in the HD+MSL renormalization prescription. In particular, the $\beta$-function of ${\cal N}=1$ SQCD can be expressed in terms of the Adler $D$-function. If all favors have the same absolute value of the electromagnetic charges, it is also possible to write a simple relation between the $\beta$-functions for the strong and electromagnetic coupling constants. In this particular case there is a special  renormalization group invariant relation.

\end{abstract}

\vspace*{-13.2cm}

\begin{flushright}
INR-TH-2024-010
\end{flushright}

\vspace*{12.2cm}

\section{Introduction}
\hspace*{\parindent}

The renormalization group method \cite{StueckelbergdeBreidenbach:1952pwl,Gell-Mann:1954yli,Bogolyubov:1956gh} plays an important role in physical considerations. For example its
enables to study  the problems  of   scale and scheme dependence of the perturbation theory predictions for Green functions and Adler $D$-function \cite{Adler:1974gd} in particular. These studies 
allow 
to investigate the consequences of  taking into account   different strong interactions contributions into various inclusive processes  generated by  the  the one-photon
 $e^+e^-$--annihilation.  Another important outcome of the renormalization group applications   is  the indication to the  unification of running couplings, imporatnt from the point of view of   testing various consequences of the Grand Unified theories  and , in particular, of  the supersymmetric ones   \cite{Salam:1978jz,Mohapatra:1986uf,Ellis:1990wk ,Amaldi:1991cn,Langacker:1991an}. 

In general, 
the latter   theories are  interesting from the theoretical point of view due to numerous attractive features of  their ultraviolet divergences \cite{West:1990tg,Buchbinder:1998qv}. 
Even in ${\cal N}=1$ supersymmetric gauge theories, in addition to the nonrenormalization of the superpotential \cite{Grisaru:1979wc}, there are some statements which can be considered as nonrenormalizations theorems.  Their applications  are leading to theoretical supports of the property of the relation of the corresponding  gauge $\beta$-function 
 to the anomalous dimension of the matter superfields by the Novikov, Shifman, Vainshtein and Zakharov (NSVZ) equation \cite{Novikov:1983uc,Jones:1983ip,Novikov:1985rd,Shifman:1986zi}.

Although this equation is  not valid in the formulated in  $D$-dimensions $\overline{\mbox{DR}}$ scheme \cite{Jack:1996vg,Jack:1996cn,Jack:1998uj,Harlander:2006xq}  (see \cite{Mihaila:2013wma} for review),
it turned out that the all-order NSVZ schemes,  in which the corresponding strict relations  are restored on the renormalized level 
after the  considered in Refs.\cite{Jones:1983ip,Kazakov:1984bj,Kazakov:1984bh} finite renormalizations ,  
can be constructed  using the typical to four-dimesions  Slavnov's higher covariant derivative (HD) regularization \cite{Slavnov:1971aw,Slavnov:1972sq,Slavnov:1977zf} and  the procedure of  minimal subtraction of logarithms (MSL) \cite{Kataev:2013eta,Shakhmanov:2017wji}.
In the formulated in Refs. \cite{Kataev:2013eta,Shakhmanov:2017wji}
 resulting HD+MSL scheme a theory is regularized by higher covariant derivatives and only powers of $\ln\Lambda/\mu$ (where $\Lambda$ is the ultraviolet cut-off and $\mu$ is renormalization point) are included into the renormalization constants. The whole class of the NSVZ schemes was described in \cite{Goriachuk:2018cac,Goriachuk:2019aqy,Goriachuk:2020wyn}.

The NSVZ-like equation has also been constructed and proved in all orders for the Adler $D$-function  in ${\cal N}=1$ supersymmetric chromodynamics (SQCD) interacting with the supersymmetric electromagnetic field  \cite{Shifman:2014cya,Shifman:2015doa}. It looks similar to the similar  equation for ${\cal N}=1$ supersymmetric electrodynamics (SQED) \cite{Vainshtein:1986ja,Shifman:1985fi} and is closely related to the NSVZ $\beta$-functions of the considered on the renormalized level in Ref.\cite{Kataev:2014gxa,Stepanyantz:2020uke}
non-Abelian Yang-Mills theory and 
for the  theories with multiple gauge couplings \cite{Shifman:1996iy,Korneev:2021zdz,
Haneychuk:2022qvu}, which are applicable even for Minimal Supersymmetric Standard Model .

In this paper we note that the resemblance between the NSVZ equations for the $\beta$-functions and the Adler $D$-function in ${\cal N}=1$ SQCD implies that the renormalization group running of the strong and electromagnetic couplings of the combined SQED+SQCD model  are related by the  certain exact equations. 

The paper is organized as follows. In Sect. \ref{Section_SQCD+SQED} we consider the supersymmetric generalization of QCD interacting with the supersymmetric  electromagnetic field assuming that the electromagnetic charges are the same for all flavors. This theory is invariant under the transformations of the group $G\times U(1)$ and contains two gauge coupling constants. Starting from the NSVZ equations for them, we find the all-loop relation between these coupling constants and construct from them a renormalization group invariant  expression. A similar theory containing the matter superfields with different $U(1)$ charges is investigated in Sect. \ref{Section_Different_Charges}. In this case it is possible to relate the ${\cal N}=1$ SQCD $\beta$-function (which is obtained in the limit of vanishing electromagnetic coupling constant) to the Adler $D$-function.

\section{Relation between gauge couplings in ${\cal N}=1$ SQCD+SQED related model}
\hspace*{\parindent}\label{Section_SQCD+SQED}

Let us investigate a massless supersymmetric QCD-like theory interacting with the Abelian gauge field in a supersymmetric way. The theory under consideration,  namely SQCD+SQED, 
 is invariant under the gauge group $G\times U(1)$, where $G$ is a simple group generalizing $SU(3)$ for usual QCD, and $U(1)$ describes the supersymmetric 
analog of the electromagnetic interaction. The corresponding coupling constants will be denoted by $g$ and $e$, respectively. Assuming that there are $N_f$ flavors (numerated by the subscript $\mbox{a}$) {\it with the same electromagnetic charges}, the superfield action of the theory can be written in the form

\begin{eqnarray}\label{Action_SQCD+SQED}
&& S = \frac{1}{2g^2}\,\mbox{Re}\,\mbox{tr}\int d^4x\,d^2\theta\, W^a W_a + \frac{1}{4e^2}\,\mbox{Re}\int d^4x\, d^2\theta\, \bm{W}^a \bm{W}_a\nonumber\\
&&\qquad\qquad\qquad\qquad\quad + \sum\limits_{\mbox{\scriptsize a}=1}^{N_f}\, \frac{1}{4}\int d^4x\, d^4\theta\, \Big(\phi_{\mbox{\scriptsize a}}^+ e^{2V + 2\bm{V}}\phi_{\mbox{\scriptsize a}}
+ \widetilde\phi_{\mbox{\scriptsize a}}^+ e^{-2V^T - 2\bm{V}} \widetilde\phi_{\mbox{\scriptsize a}}\Big).\qquad
\end{eqnarray}

\noindent
Here $V$ and $\bm{V}$ are the gauge superfields corresponding to the groups $G$ and $U(1)$, respectively. The chiral matter superfields $\phi_{\mbox{\scriptsize a}}$ and $\widetilde\phi_{\mbox{\scriptsize a}}$ belong to the (conjugated) representations $R$ and $\xbar R$ of G and have the opposite $U(1)$ charges. In the gauge part of the action $V = gV^A t^A$, where $t^A$ are the generators of the fundamental representation of the group $G$, which satisfy the normalization condition $\mbox{tr}(t^A t^B) = \delta^{AB}/2$. In the term which contains matter superfields $V = gV^A T^A$, where $T^A$ are the generators of the representation $R$. Two supersymmetric gauge superfield strengths for the considered theory are written in the form

\begin{equation}
W_a = \frac{1}{8} \xbar D^2\Big(e^{-2V} D_a e^{2V}\Big);\qquad \bm{W}_a = \frac{1}{4} \xbar D^2 D_a \bm{V}.
\end{equation}

\noindent
For the SQCD and SQED coupling constants we will also use the notations $\alpha_s\equiv g^2/4\pi$ and $\alpha\equiv e^2/4\pi$. The corresponding $\beta$-functions are defined as

\begin{equation}\label{Betas_Definition}
\beta_s(\alpha_s,\alpha) \equiv \frac{d\alpha_s}{d\ln\mu}\bigg|_{\alpha_0,\alpha_{s0}=\mbox{\scriptsize const}};\qquad
\beta(\alpha,\alpha_s) \equiv \frac{d\alpha}{d\ln\mu}\bigg|_{\alpha_0,\alpha_{s0}=\mbox{\scriptsize const}},
\end{equation}

\noindent
where the differentiation is made at fixed values of the bare constants denoted by $\alpha_0$ and $\alpha_{s0}$. 

The chiral matter superfields  $\phi_{\mbox{\scriptsize a}}$ and $\widetilde\phi_{\mbox{\scriptsize a}}$ are renormalized as 

\begin{equation} 
\label{Zconstant} 
\phi_{\mbox{\scriptsize a}}=Z_a(\alpha_s,\alpha)^{1/2}\phi_{\mbox{\scriptsize a,0}}
;\qquad \widetilde\phi_{\mbox{\scriptsize a}}
 = Z_a(\alpha_s,\alpha)^{1/2} \widetilde\phi_{\mbox{\scriptsize a,0}}, 
\end{equation}
where the unrenormalized bare fields are marked by extra index 0.  The related anomalous dimension of the matter superfields reads as 

\begin{equation}
\label{AD}
\gamma_a(\alpha_s,\alpha) \equiv \frac{d ln(Z_a)}{d\ln\mu}\bigg|_{\alpha_0,\alpha_{s0}=\mbox{\scriptsize const}}
\end{equation}

The theory investigated in this section is obtained from the one considered in \cite{Korneev:2021zdz} after setting $q_{\mbox{\scriptsize a}} = 1$ for all $\mbox{a}$. In this case all anomalous dimensional of the chiral matter superfields for the theory (\ref{Action_SQCD+SQED}) evidently coincide, namely

\begin{equation}
\label{ADex}
\gamma_a(\alpha_s,\alpha)_I{}^J = \gamma(\alpha_s,\alpha)\cdot \delta_I^J,
\end{equation}

\noindent
where $I$ and $J$ include both the indices numerating chiral matter superfields $\phi_{\mbox{\scriptsize a}}$ and $\widetilde\phi_{\mbox{\scriptsize a}}$ and  correspond  to the representations $R$ or $\xbar R$ of the coloured  group .

The corresponding  equations for the $\beta$-functions of  both coupling constants in the model  under consideration have been constructed in \cite{Korneev:2021zdz} and read :

\begin{eqnarray}\label{Beta_G}
&& \frac{\beta_s(\alpha_s,\alpha)}{\alpha_s^2}\bigg|_{\rm{NSVZ}} = - \frac{1}{2\pi(1-C_{2} \alpha_s/2\pi)} \bigg[\, 3 C_2 - 2 T(R) N_f \Big(1-\gamma(\alpha_s,\alpha)\Big) \bigg];\qquad\\
\label{Beta_U(1)}
&& \frac{\beta(\alpha,\alpha_s)}{\alpha^2}\bigg|_{\rm{NSVZ}} = \frac{1}{\pi}\, \mbox{dim}\,R\, N_f \Big(1-\gamma(\alpha_s,\alpha)\Big),
\end{eqnarray}

\noindent
where $\mbox{dim}\,R$ is the dimension of the representation $R$, and we use the notations

\begin{eqnarray}
&& f^{ACD} f^{BCD} = C_2 \delta^{AB};\qquad \mbox{tr}(T^A T^B) = T(R)\delta^{AB};\qquad\vphantom{\Big(}\nonumber\\
&& [t^A,t^B] = if^{ABC} t^C;\qquad\quad  [T^A,T^B] = if^{ABC} T^C.\vphantom{\Big(}
\end{eqnarray}

Comparing the equations (\ref{Beta_G}) and (\ref{Beta_U(1)}) we conclude that the anomalous dimensions of the matter superfields can be excluded, so that the $\beta$-functions for the non-Abelian and Abelian supersymmetric coupling constants satisfy the following  all-order exact equation

\begin{equation}\label{Exact_Beta_Relation}
\Big(1-\frac{C_{2}\alpha_s}{2\pi}\Big) \frac{\beta_s(\alpha_s,\alpha)}{\alpha_s^2}\bigg|_{\rm{NSVZ}} = - \frac{3 C_2}{2\pi} + \frac{T(R)}{\mbox{dim}\,R}\cdot \frac{\beta(\alpha,\alpha_s)}{\alpha^2}\bigg|_{\rm{NSVZ}}.
\end{equation}

\noindent
According to \cite{Stepanyantz:2011jy} (see also \cite{Smilga:2004zr} for   the Abelian case)  the  NSVZ-like equations are valid for both  renormalization group functions  defined in terms of the bare couplings if a theory is regularized by higher covariant derivatives independently on a way in which divergences are removed. 
In the renormalized case this implies that these functions  are defined  in the HD+MSL scheme.\footnote{As shown in Ref.\cite{Smilga:2004zr} at two-loop level and in Ref. \cite{Kataev:2019olb} in all loops,  in 
 the Abelian case the NSVZ equation is also satisfied in the on-shell scheme.} Therefore, Eq. (\ref{Exact_Beta_Relation}) is valid in all orders in the case of applying this HD+MSL  renormalization prescription.

Using Eq. (\ref{Betas_Definition}) and integrating Eq. (\ref{Exact_Beta_Relation}) over $\ln\mu$ we obtain the relation between the strong and electromagnetic coupling constants in the theory under consideration,

\begin{eqnarray}\label{Integrated_Relation}
&& \frac{1}{\alpha_s} - \frac{1}{\alpha_{s0}} + \frac{C_2}{2\pi} \ln\frac{\alpha_s}{\alpha_{0s}} = - \frac{3C_2}{2\pi}\ln\frac{\Lambda}{\mu} + \frac{T(R)}{\mbox{dim}\,R}\Big(\frac{1}{\alpha} - \frac{1}{\alpha_0}\Big) + C.
\end{eqnarray}

\noindent
Here the cut-off   $\Lambda$ is the analog of $\Lambda_{QCD}$-parameter and the constant  $C$ is  analogous to the one introduced in \cite{Vladimirov:1979ib,Vladimirov:1979my} and is 
 fixing the scale-dependence of the ratio $\Lambda/\mu$. 

 In the HD+MSL scheme the coupling constants satisfy the boundary conditions 

\begin{equation}
\alpha_s\Big(\alpha_{s0},\alpha_0,\ln\frac{\Lambda}{\mu}\Big)\bigg|_{\mu=\Lambda} = \alpha_{s0};\qquad \alpha\Big(\alpha_{0},\alpha_{s0},\ln\frac{\Lambda}{\mu}\Big)\bigg|_{\mu=\Lambda} = \alpha_{0},
\end{equation}

\noindent
so that in this scheme $C=0$. 
Keeping in mind the the  considerations of Refs.\cite{Hisano:1997ua, Avdeev:1997vx} and 
taking into account that the bare couplings do not depend on the renormalization scale $\mu$   
we  rewrite Eq. (\ref{Integrated_Relation}) in the following exactly  renormalization group invariant form 
\begin{equation}\label{RGI}
\Big(\alpha_{s0}
\Big)^{C_2} \exp\Big(\frac{2\pi}{\alpha_{s0}} - \frac{T(R)}{\mbox{dim}\,R}\cdot \frac{2\pi}{\alpha_0}\Big) =
\Big(\alpha_s \Big)^{C_2} \exp\Big(\frac{2\pi}{\alpha_s} - \frac{T(R)}{\mbox{dim}\,R}\cdot \frac{2\pi}{\alpha}+3C_2\ln\frac{\Lambda}{\mu}
-2\pi C \Big)= \mbox{RGI} ~~.
\end{equation}

\noindent
Here  RGI is short for the renormalization group invariant, i.e. the expression which vanishes after differentiating with respect to $\ln\mu$ (at fixed values of $\Lambda$ and the bare couplings).

Note that Eq. (\ref{RGI}) is valid only for certain renormalization prescriptions , which within the ways of studies of Ref.\cite{Goriachuk:2018cac} and Refs.\cite{Goriachuk:2019aqy,Goriachuk:2020wyn} 
are forming  the {\it class} or the {\it subgroup} of the NSVZ-type renormalization presriprtions.  Togerther with the HD+MSL presription of Refs.\cite{Kataev:2013eta,Shakhmanov:2017wji}
  this subgroups includes the 
considered in Ref. \cite{Kataev:2019olb}
 on-shell presrription, but not  the $\overline{\mbox{DR}}$ -scheme, which violates the validity of the NSVZ-type SUSY relations on the renormalized level .

\section{The general ${\cal N}=1$ SQCD+SQED  case}
\hspace*{\parindent}\label{Section_Different_Charges}

 Consider now  the theory analysed  in the previous section but with  the matter superfields of  different $U(1)$ charges.  The corresponding action is written now  in the form

\begin{eqnarray}\label{Action_Generalized}
&& S = \frac{1}{2g^2}\,\mbox{Re}\,\mbox{tr}\int d^4x\,d^2\theta\, W^a W_a + \frac{1}{4e^2}\,\mbox{Re}\int d^4x\, d^2\theta\, \bm{W}^a \bm{W}_a\nonumber\\
&&\qquad\qquad\qquad\qquad\quad + \sum\limits_{\mbox{\scriptsize a}=1}^{N_f}\, \frac{1}{4}\int d^4x\, d^4\theta\, \Big(\phi_{\mbox{\scriptsize a}}^+ e^{2V + 2q_{\mbox{\scriptsize a}} \bm{V}}\phi_{\mbox{\scriptsize a}}
+ \widetilde\phi_{\mbox{\scriptsize a}}^+ e^{-2V^T - 2q_{\mbox{\scriptsize a}} \bm{V}} \widetilde\phi_{\mbox{\scriptsize a}}\Big).\qquad
\end{eqnarray}

\noindent
This expression differs from the one in Eq. (\ref{Action_SQCD+SQED}) in that the Abelian gauge superfield in the matter terms is multiplied by their charges  $q_{\mbox{\scriptsize a}}$.
 In this case the  SQCD  $\beta_s$-function  
and the defined in  Eq.(\ref{Betas_Definition}) 
anomalous dimensions of the matter superfields   will depend now   from these charges. For example, the right hand side of Eq.(\ref{ADex})  for 
 the anomalous dimensions of $\phi_{\mbox{\scriptsize a}}$ and $\widetilde\phi_{\mbox{\scriptsize a}}$   contain now the  $\mbox{a}$-dependence : 

\begin{equation}
\gamma_{\mbox{\scriptsize a}}(\alpha_s,\alpha)_i{}^j = \gamma_{\mbox{\scriptsize a}}(\alpha_s,a)\cdot \delta_i^j~~~.
\end{equation}

Further on we will  be 
interested in the limit $\alpha = e^2/4\pi\to 0$. In this case the renormalization group running of the strong coupling constant $\alpha_s$ is exactly the same as in usual ${\cal N}=1$ SQCD with the gauge group $G$ and $N_f$ flavors (which, in particular, contains $N_f$ Dirac fermions in the irreducible representation $R$). The running of the electromagnetic coupling constant due to taking into account 
strong interactions effects 
 is described by the Adler $D$-function \cite{Adler:1974gd} which is related to the $\beta$-function of  the coupling constant $\alpha$ in the limit $\alpha\to 0$ as 

\begin{equation}\label{Adler_Definition}
D(\alpha_s)\equiv -\frac{3\pi}{2}\frac{d}{d\ln\mu}\Big(\frac{1}{\alpha}\Big)\bigg|_{\alpha_0,\alpha_{s0}=\mbox{\scriptsize const},\ \alpha\to 0} = \frac{3\pi}{2} \lim\limits_{\alpha\to 0} \frac{\beta(\alpha,\alpha_s)}{\alpha^2}.
\end{equation}

\noindent
In the limit $\alpha\to 0$ the anomalous dimensions of the matter superfields do not depend on $\alpha$ and, therefore, on $q_{\mbox{\scriptsize a}}$. This implies that in this case all anomalous dimensions of chiral matter superfields are the same, namely 

\begin{equation}
\lim\limits_{\alpha\to 0}\gamma_{\mbox{\scriptsize a}}(\alpha_s,\alpha) = \gamma(\alpha_s)~~~.
\end{equation}

\noindent

The exact NSVZ-like expression for the Adler $D$-function in the theory under consideration has been derived in \cite{Shifman:2014cya,Shifman:2015doa} 
 (see also \cite{Kataev:2017qvk}) and is written as

\begin{equation}\label{Alder_Exact}
D(\alpha_s)\Big|_{\rm{NSVZ}} =  \frac{3}{2}\, \mbox{dim}\,R \sum\limits_{\mbox{\scriptsize a}=1}^{N_f} q_{\mbox{\scriptsize a}}^2\Big(1- \gamma(\alpha_s)\Big)\equiv \frac{3}{2}\,\bm{q^2}\, \mbox{dim}\,R \Big(1- \gamma(\alpha_s)\Big),
\end{equation}

\noindent
where we introduced the notation

\begin{equation}
\bm{q^2} \equiv \sum\limits_{\mbox{\scriptsize a}=1}^{N_f} q_{\mbox{\scriptsize a}}^2.
\end{equation}

Comparing now Eq. (\ref{Alder_Exact}) with Eq.(\ref{Beta_G}) as given in the previouse Section and taking the limit  $\alpha\to 0$
we conclude  that  
the NSVZ $\beta$-function of ${\cal N}=1$ SQCD can be expressed in terms of the Adler $D$-function by the following  all-loop equation as 

\begin{equation}\label{Exact_Beta_Adler_Relation}
\beta_s(\alpha_s)\Big|_{\rm{NSVZ}} = \lim\limits_{\alpha\to 0}\beta_s(\alpha_s,\alpha)\Big|_{\rm{NSVZ}}=
 - \frac{\alpha_s^2}{2\pi(1-C_{2} \alpha_s/2\pi)} \bigg[\, 3 C_2 - \frac{4\, T(R) N_f D(\alpha_s)\Big|_{\rm{NSVZ}}}{3\,\bm{q^2}\, \mbox{dim}\,R }  \bigg]~~.
\end{equation}

\noindent
It results from the consideration of the  renormalization group running of the strong and eletromagnetic coupling constants in the theory (\ref{Action_Generalized}) in the limit $\alpha\to 0$. This equation is 
valid for the bare unrenormalizaed coupling constant $\alpha_s=\alpha_{s0}$ and for the  renormalized one $\alpha_{s}$ , fixed within ${\it subgroup}$ of the  in the NSVZ-type renormalization schemes and 
thus HD+MSL scheme in all orders of pertrurbation theory.

Integrating Eq. (\ref{Exact_Beta_Adler_Relation}) over $\ln\mu$ we obtain the equation anlogous to Eq. (\ref{Integrated_Relation}) , 

\begin{eqnarray}\label{Integrated_Relation_Adler}
&& \lim\limits_{\alpha\to 0}\bigg[ \frac{1}{\alpha_s} - \frac{1}{\alpha_{s0}} + \frac{C_2}{2\pi} \ln\frac{\alpha_s}{\alpha_{0s}} - \frac{  T(R)N_f }{\bm{q^2} \mbox{dim}\,R} \Big(\frac{1}{\alpha} - \frac{1}{\alpha_0}\Big)\bigg] = - \frac{3C_2}{2\pi}\ln\frac{\Lambda}{\mu} + C.
\end{eqnarray}

\noindent
where for the NSVZ-type renormalization schemes we have   $C=0$ .  It may be of interset to consider possible  consequencies of application of the solution of this 
exact equatiion say in  the SUSY Grand Unified model considerations.

\section{Conclusion}
\hspace*{\parindent}

We demonstrated that in ${\cal N}=1$ SQCD interacting with ${\cal N}=1$ SQED there are the exact all-loop equations relating the renormalization group running of the strong and electromagnetic coupling constants. Namely, if for all matter superfields absolute values of electromagnetic charges are equal, then it is possible to relate two $\beta$-functions of theory by Eq. (\ref{Exact_Beta_Relation}). This equation implies to the existence of the renormalization group invariant (\ref{RGI}). If the (absolute values of the) electromagnetic charges of matter superfields are different, then it is possible to construct the equation (\ref{Exact_Beta_Adler_Relation}) which relates the $\beta$-function of ${\cal N}=1$ SQCD to the Adler $D$-function.

All exact equations constructed in this paper are valid in the HD+MSL schemes, when the theory is regularized by higher derivatives, and divergences are removed by minimal subtractions of logarithms. 
 In the  $\overline{\mbox{DR}}$-scheme  the relation (\ref{Exact_Beta_Adler_Relation}) is not satisfied starting from the three-loop approximation, where the scheme dependence becomes essential. The similar feature will probably demonstrate 
itself in the applied within lattice studies of Ref. \cite{Bonanno:2024bqg} renormalization 
procedure, which is based on not retaining supersymmetry $\overline{\mbox{DR}}$-scheme, but the  $\overline{\mbox{MS}}$-one.

Finally, let us mention that, although we do not expect the existence of exact equations relating the running of the gauge couplings in usual QCD, some features of supersymmetric theories may nevertheless be retained even in the non-supersymmetric case.

\section{Acknowledgements} 
The results  related to this work were  first reported  at  Fradkin Centennial Conference ( Lebedev Institute, Moscow, September 2-6, 2024). We are grateful to 
A.  V. Smilga, M. A. Vasiliev and A. A. Tseytlin for the interest, questions and comments.

\end{document}